# Attraction by pairwise coherence explains the emergence of ideological sorting

Federico Zimmerman [a,b,c,d,e,f,1,]*, Lucía Pedraza [d,g,1], Joaquín Navajas [a,b,c], Pablo Balenzuela [d,g]

a. Laboratorio de Neurociencia, Universidad Torcuato Di Tella, Buenos Aires, Argentina
b. Consejo Nacional de Investigaciones Científicas y Técnicas (CONICET), Buenos Aires, Argentina
c. Escuela de Negocios, Universidad Torcuato Di Tella, Buenos Aires, Argentina
d. Universidad de Buenos Aires, Facultad de Ciencias Exactas y Naturales, Departamento de Física, Ciudad Universitaria, Buenos Aires, Argentina.
e. Harvard University, Harvard Business School, Boston, USA
f. Harvard University, Digital, Data and Design Institute, Cambridge, USA
g. CONICET - Instituto de Física Interdisciplinaria y Aplicada (INFINA), Ciudad Universitaria, Buenos Aires, Argentina.

1. F.Z and L.P. contributed equally to this work.
* To whom correspondence may be addressed. **Email**: fzimmerman@hbs.edu

# Attraction by pairwise coherence explains the emergence of ideological sorting

**Abstract**

Political polarization has become a growing concern in democratic societies, as it drives tribal alignments and erodes civic deliberation among citizens. Given its prevalence across different countries, previous research has sought to understand under which conditions people tend to endorse extreme opinions. However, in polarized contexts, citizens not only adopt more extreme views but also become correlated across issues that are, a priori, seemingly unrelated. This phenomenon, known as "ideological sorting", has been receiving greater attention in recent years but the micro-level mechanisms underlying its emergence remain poorly understood. Here, we study the conditions under which a social dynamic system is expected to become ideologically sorted as a function of the mechanisms of interaction between its individuals. To this end, we developed and analyzed a multidimensional agent-based model that incorporates two mechanisms: homophily (where people tend to interact with those holding similar opinions) and pairwise-coherence favoritism (where people tend to interact with ingroups holding politically coherent opinions). We numerically integrated the model's master equations that perfectly describe the system's dynamics and found that ideological sorting only emerges in models that include pairwise-coherence favoritism. We then compared the model's outcomes with empirical data from 24,035 opinions across 67 topics and found that pairwise-coherence favoritism is significantly present in datasets that measure political attitudes but absent across topics not considered related to politics. Overall, this work combines theoretical approaches from system dynamics with model-based analyses of empirical data to uncover a potential mechanism underlying the pervasiveness of ideological sorting.

**Significance statement**

We investigate the mechanisms behind ideological sorting, a phenomenon in which people's opinions become aligned on seemingly unrelated topics. By implementing a multidimensional agent-based model that includes only experimentally validated psychological phenomena such as homophily (the tendency for people to interact with those who share similar opinions) and pairwise-coherence favoritism (the tendency for people to interact with ingroups that hold politically coherent opinions), we found that ideological sorting is primarily driven by the latter. Moreover, we support our findings by linking the model to empirical data, revealing that the influence of pairwise-coherence favoritism is present in political attitudes but absent in issues not considered related to politics.

**MAIN TEXT**

**INTRODUCTION**

The increasing political polarization (1–3) has become a worrying concern in many different countries (4) and a serious threat to society and democracy itself (5). Polarization drives hatred among family members (6), enables the spread of misinformation (7, 8), promotes the segregation of societies, and reduces the chances of coherently responding to large-scale crises, as recently demonstrated by the COVID-19 pandemic (9–11). Concerned about the risks and societal impacts of this phenomenon, researchers and policymakers have tried to develop interventions to reduce polarization, obtaining mixed results and demonstrating the complexity of the problem (12–14). In this context, understanding why societies tend to become more polarized has become a crucial issue in the behavioral and social sciences.

One promising way to understand the emergence of political polarization is by studying the behavior of agent-based models (ABMs) under different conditions of social influence and interactions. From a modeling point of view, several mechanisms have been explored in order to explain issue polarization. For instance, bounded confidence (15–18), negative influence or repulsion (19–21), or homophily in conjunction with other mechanisms such as social influence in cultural vectors (22–25), persuasive arguments theory (26, 27), or biased assimilation (28). Based on a few simplistic assumptions, most studies have focused on understanding under which conditions individuals may polarize and become more extreme on one single topic (29), even though this simplification lacks the possibility of modeling phenomena that arise due to the interplay of multiple issues.

While disagreement on policy issues has been extensively studied in previous research, relatively less attention has been paid to understanding why people tend to be more aligned across diverse and seemingly unrelated topics (2, 30, 31). For example, an individual who supports the women's right to voluntarily terminate pregnancy will be more likely to support stricter legislation on gun control, even though these topics are, a priori, unrelated to each other. While there is consensus on the existence and importance of this phenomenon, known as "ideological sorting", there is debate about whether it has been increasing in recent years (32–35). In any case, the micro-level mechanisms underlying the emergence of ideological sorting in social systems remain poorly understood.

In a recent paper, a set of large-scale behavioral experiments have shown that people not only hold politically coherent opinions across very different issues but also that this property, i.e., political coherence, increases interpersonal attraction among co-partisans (36). In other words, individuals who hold coherent opinions are more attractive than those individuals having some degree of ambivalence in their attitudes (e.g., a person who is anti-abortion but supports gun control). This idea is in line with previous findings showing that people favor pro-norm deviants (37, 38). However, whether and how this driver of interpersonal attraction, called "pairwise-coherence favoritism", relates to macro-level patterns of political polarization and partisan-ideological sorting remains largely unknown.

This overreaching aim necessarily requires the formulation of multidimensional dynamic models where polarization could arise in independent individual topics (with no correlation between them) or as ideological states where topics are aligned and correlations between them are pronounced. In proportion, there are far fewer models that study opinion in multidimensional spaces, which could give rise to these phenomena (39–44). Ideological sorting has been previously modeled by considering continuous opinions and non-orthogonal and overlapping topics (45) or directional voting (46). However, none of these approaches have tested the effect of pairwise-coherence favoritism, given that it is a recently uncovered empirical finding in the social sciences. Here we show that by incorporating this rule of interpersonal attraction, issue alignment emerges, even when agents start from a random distribution of opinions. To demonstrate this, we first formulated the model, numerically integrated its master equations, and ran multiple computational simulations, always obtaining the same consistent results. We then compared different final states with actual data from multiple datasets that include 24,035 opinions on different controversial issues. All analyses indicate that homophily alone is insufficient to account for ideologically sorted states, highlighting the significance and impact of pairwise-coherence favoritism in political interactions.

**THE MODEL**

In previous research (36), the authors performed three different studies in different countries and found that people are more attracted to politically coherent ingroups rather than to those who hold ambiguous or ambivalent opinions. In that study, someone was considered coherent if her/his opinions were aligned with her/his political ideology. For example, a coherent Democrat would be pro-choice regarding abortion and also favor stricter gun control. In two live crowd experiments, participants were arranged in dyads, discussed five controversial topics, and completed an interpersonal attraction questionnaire. In both cases, interpersonal liking increased as a function of similarity, but also of pairwise coherence. Interpersonal liking was found to be nonreciprocal: people with ambivalent and uncertain political views were more attracted to coherent ingroups than vice versa. These results were validated by performing an online preregistered experiment where political coherence was experimentally manipulated. Overall, these empirical results suggested that liking in the political domain may not be solely driven by homophily but by more complex notions of group affiliation such as pairwise coherence.

**Agents, opinions, and communities**

We considered a system of N agents. Each agent holds a multidimensional vector opinion, where each dimension stands for the agent's opinion on a particular issue (for the sake of simplicity, we considered only two issues). For each issue, the agent could be against, undecided, or in favor of the issue. Two considerations about how opinions are modeled are to be made: Firstly, the inclusion of a neutral or undecided state is grounded on modeling approaches (47–50), behavioral experiments (48, 51), and surveys (e.g., ANES). Secondly, in order to define a metric for quantifying pairwise coherence, we assumed that every political issue expressing a right-wing opinion will be labeled as +1, a left-wing opinion as -1, and undecided as 0. Additionally, we explored the implications of agents having more than three possible opinions. For example, when considering a scenario with five possible opinions, we found that the results are the same as when considering only three possible opinions, though the model becomes more complex (see appendix for details).

Therefore, in terms of coherence, we can define four different communities: Coherent agents (C): Agents that hold assertive and matching opinions on both issues. Left-wing coherent agents ($C_L$) hold both left-wing opinions (O=(-1,-1)) and right-wing coherent agents ($C_R$) hold right-wing opinions (O=(1,1)). Incoherent agents (I): Agents that hold assertive but opposite opinions on the two different issues. Incoherent agents hold one left-wing opinion and one right-wing opinion (O=(-1,1) or O=(1,-1)). Weak agents (W): Agents that hold an assertive opinion on one issue and are undecided regarding the other one. The agent's ideology is determined by the political leaning of its assertive opinion (O=(-1,0) or O=(0,-1) are considered left-wing weak agents and O=(1,0) or O=(0,1) right-wing weak agents). Apathetic agents (A): Agents that are undecided on both issues (O=(0,0)). Following these definitions, every possible agent's opinions can be mapped onto a 3x3 board as shown in Fig. 1A.

In this work, we focused on the populations' dynamics of each community. The agents' opinions were initially independent and uniformly distributed. Because the number of possible combinations of opinions is not the same for each community, the initial proportion of agents for each community varies. For example, the initial proportion of coherent and incoherent agents is 2/9, while the proportion of weak agents is 4/9.

**Definitions of similarity and pairwise coherence**

In what follows, we show how we defined similarity and pairwise coherence and how to implement them in the interaction mechanism between agents. First, we defined similarity between the agents *i* and *j* ($S_{ij}$) as a function of the Manhattan distance between the agents' opinions (i.e., the sum of the absolute differences of their Cartesian coordinates). If two agents hold the same two opinions, similarity is 1, and it is 0 if they have opposite stances on both issues. This is computed as:

$$S_{ij} = 1 - \frac{|(O_i - O_j)|_1}{4}. \quad [1]$$

For example, as depicted in Fig. 1B, for agents *i* and *j* with opinions $O_i=(0,1)$ and $O_j=(1,1)$, the similarity between them is $S_{ij}=S_{ji}=3/4$, while for agent *m* with opinion $O_k=(-1,0)$, the similarity with *i* is $S_{ik}=S_{ki}=2/4$.

Second, in order to implement attraction from pairwise coherence, we define two metrics.
The agents' ideology is computed as:

$$I_i = \frac{O_i^{(x)} + O_i^{(y)}}{2}, \quad [2]$$

which ranges from -1 to 1. $I_i$=-1 corresponds to left-wing coherent agents holding two left-wing opinions and $I_i$=+1 corresponds to right-wing ones. Weak agents have an absolute value of $I_i$=0.5 and incoherent agents $I_i$=0. The ideology's sign value corresponds to the agent's leaning: a positive value describes right-wing agents and a negative one describes left-wing agents. Agents whose ideology is 0 are neutral as they do not belong to any of the two groups. Pairwise coherence is computed as:

$$C_{ij} = |I_i|\delta ij \quad [3]$$

(where δ is 1 if *i* and *j* have the same ideology's sign and 0 otherwise), depends on both agents' ideology and the partner agent's coherence. It takes positive values for dyads who share the same leaning and is 0 otherwise. Its maximum possible value corresponds to coherent ingroup agents and it is 0 for outgroup agents. This measure is not commutative and this is in line with experimental results that showed that social influence is not always reciprocal (52).

In Fig. 1C, we show an example involving agents *i*, *j,* and *m.* The pairwise coherence that *j* perceives from *i* is $C_{ij}$=0.5, while the coherence that *i* perceives from *j* is $C_{ji}$=1. Meanwhile, the pairwise coherence between *i* and *m* is 0 as they have ideologies with opposite signs. Interestingly, pairwise coherence for neutral agents is 0 with all communities.

**Interactions dynamics**

In this model, agents interact with each other and these interactions influence agents' opinions. At each time step, two agents are randomly selected and their interaction would lead one of the agents to influence the other with probability "P". We incorporated two different mechanisms that impact agents' influence: homophily (53, 54) and pairwise coherence (36, 55), as defined previously. This probability of influence is implemented as a linear combination of both mechanisms:

$$P_{ij} = (1-k)S_{ij} + kC_{ij.} \qquad [4]$$

The parameter k modulates the strength of each of the two terms and ranges from 0 to 1. When k is 0, interactions are only driven by homophily and, when it is 1, only by pairwise coherence. Any intermediate value takes into consideration both dynamics as proposed by previous experimental work.

Pairwise interactions lead to opinion changes that could occur only in one of the two issues. Following field-experiment results that showed that partisans who are taken out of their Twitter's echo-chamber became more extreme (56), this model's interactions can be attractive, repulsive, or have no effect, depending on agents' similarity. When agents are similar enough, influence will be attractive and agent *i* will move closer to *j* by changing one of its own opinions as depicted in Fig. 1D. Conversely, for dissimilar agents they can repel or ignore each other. In the case where they are ignored (non-repulsive model), the model can be thought of as an extension to 2D of a bounded confidence mechanism (15, 18, 57). In the repulsion case, agent *i* changes one of its opinions (selected at random) moving further from agent *j* and reducing their similarity as shown in Fig. 1E. It can happen that the agent cannot move any further in the selected direction and, in this particular case, no change will be made and the interaction would have no effect at all. For example, in the case of an agent holding O = (-1, -1) that cannot move further to negative values in any dimension. We set the attraction-similarity boundary, T, at similarity T=3/4. This means that for greater or equal similarity values of 3/4, interactions are attractive and they are repulsive or have no effect otherwise. By setting T=3/4, we ensured that there are no attractive interactions between agents from different leanings, thus avoiding contradictions with previous experimental findings (56). Moreover, we explored different values of T and found that alternative settings lead to final states where all agents converge to one type of population, a scenario that is not observed in actual opinion data (see appendix for details).

**Summary of interaction rules**

In summary, let N agents hold two different opinions initially randomly distributed. For each time step: 1) Two agents *i* and *j* are randomly chosen and we compute $S_{ij}$, $I_{ij}$, and $P_{ij}$ according to their opinions. 2) Agent *i* influences *j* with probability $P_{ij}$. The influence can be attractive, repulsive, or have no effect, depending on the agents' similarity. 3) The agent *i* can modify its opinion on one of the two issues.

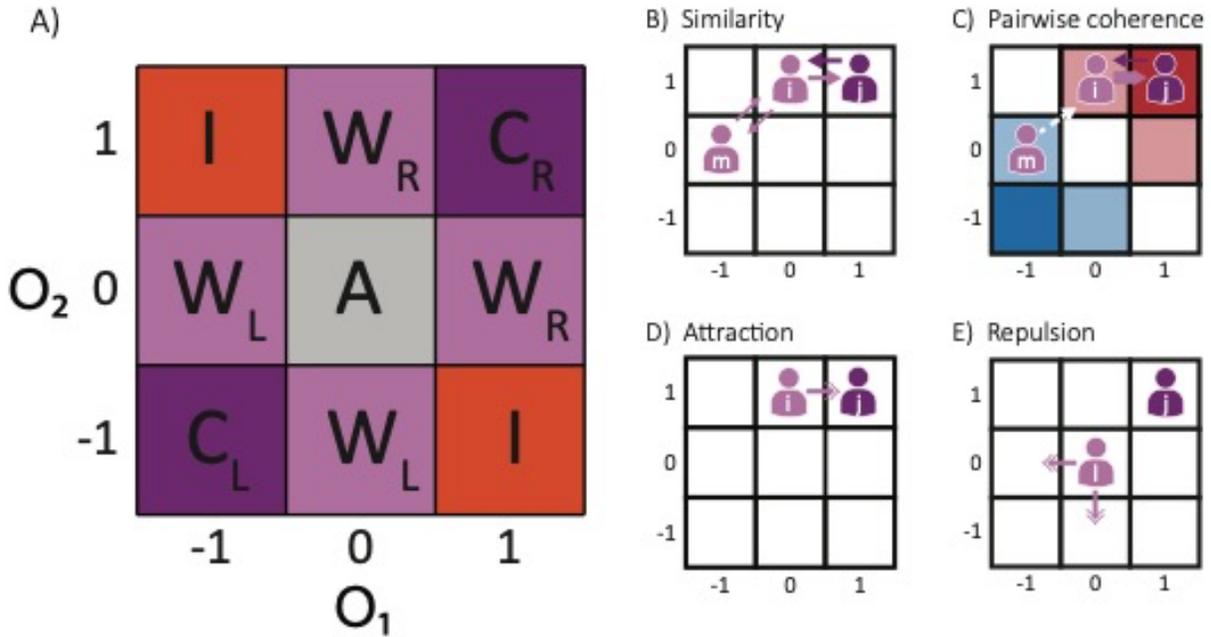

**Figure 1: The model. A)** Each agent holds an independent opinion on two different topics and, according to their opinions, they are classified in four different communities: coherent (C), incoherent (I), weak (W), or apathetic (A). Additionally, agents' opinions define their ideologies. Agents can be considered left- or right-wing (L or R) depending on whether most of their opinions are left- or right-wing oriented. **B)** Similarity measures how similar the opinions between two agents are. It is a commutative measure. Similarity between agents i and j ($S_{ij}=S_{ji}=0.75$) is higher than similarity between i and m ($S_{im}=0.5$) as the Manhattan distance is lower. **C)** Pairwise coherence measures the target agent's political coherence. This is a non-commutative measure. $C_{ji}$ is 1 because agent j is coherent, but $C_{ij}$ is 0.5 because i is a weak agent. $C_{mi}$ is 0 because the agents do not share the same ideology. While i is a right-wing oriented agent, m is left-wing oriented. **D)** Interactions between similar agents are attractive (S≥T). After an interaction between i and j, i changes its opinion and moves closer to j. **E)** Interactions between dissimilar agents are repulsive (S<T). After an interaction between l and j, l changes its opinion and moves further from j. This movement could be to the left or down with probability 0.5.

**MASTER EQUATIONS**

We developed a set of master equations that describe the dynamics of the agents' communities. As at each interaction two agents are randomly selected, the probability of choosing an agent from a particular community depends on the community's proportion. We computed the likelihood of changing opinions after pairwise interactions and obtained the flux's expected values between populations. For example, the probability of a weak agent becoming coherent or vice versa was calculated by considering the four possible scenarios in which this could occur:

- A weak agent interacts with a coherent neighbor with probability WC/2 (where W and C are the proportions of Weak and Coherent agents, respectively). With influence probability given by $P = (1-k)\frac{3}{4} + k$, the weak agent would move to the coherent population.
- A weak agent interacts with another weak agent with an opposite ideology with probability $W^2/4$. With probability $P = (1-k)\frac{1}{2}$, it would become coherent due to a repulsive interaction.
- A weak agent interacts with an opposite ideology coherent agent with probability WC/2. With probability $P = (1-k)\frac{1}{4}$, it would move to the coherent population due to a repulsive interaction.
- A coherent agent interacts with an adjacent weak agent with probability WC/2. With probability $P = (1-k)\frac{3}{4} + k/2$, the coherent agent would move to the weak population.

By repeating this analysis with the other communities, we obtained the following equations:

$$\frac{dC}{dt} = WC(\frac{1-k}{16} + \frac{k}{4}) + W^2\frac{1-k}{16} \quad [5]$$

$$\frac{dI}{dt} = WI\frac{1-k}{16} + W^2\frac{1-k}{16} \quad [6]$$

$$\frac{dA}{dt} = -AC\frac{1-k}{2} - AI\frac{1-k}{2} \quad [7]$$

$$\frac{dW}{dt} = -WC(\frac{1-k}{16} + \frac{k}{4}) - WI\frac{1-k}{16} - W^2\frac{1-k}{8} + AC\frac{1-k}{2} + AI\frac{1-k}{2} \quad [8]$$

We numerically integrate these equations with agents' initial opinions randomly distributed (i.e., C(0)=I(0)=2/9, A(0)=1/9, and W(0)=4/9). Then, we observed the final distribution of each community for different values of k, as presented in Fig. 2A. Particularly, when k=0, interactions are only driven by homophily. Coherent and incoherent populations exhibit the same behavior (dC/dt=dI/dt) and, consequently, the final agents' proportions are the same for both communities (Fig. 2A top-left inset). We observe that as k increases so does the number of coherent agents (Fig. 2A top-right inset). When k=1, interactions are only driven by pairwise coherence, and, on average, apathetic and incoherent populations do not change over time (dA/dt=dI/dt=0).

We conducted an analysis of fixed points and stability for the system of equations. To do that, we set the equations equal to 0 and assumed that k≠1. If W≠0, from equation 6, we obtained that W=-I. Given that the four variables represent proportions, their values should be between 0 and 1, it is not feasible for W to be non-zero; therefore, W=0. If A≠0, from equation 7, we obtained that C=-I. Since these variables cannot take negative values, C and I must also be zero. Then, considering that C+I+A+W=1, we concluded that A=1, obtaining a fixed point at C=I=W=0 and A=1. If A=0, all the equations become null, so the set of points of the form I=1-C, W=A=0 are also fixed points. To analyze the stability at these fixed points, we linearized the system and calculated the eigenvalues of the Jacobian matrix. The Jacobian for the first fixed point is:

$$\begin{pmatrix} 0 & 0 & 0 & 0 \\ 0 & 0 & 0 & 0 \\ -\frac{1-k}{2} & -\frac{1-k}{2} & 0 & 0 \\ \frac{1-k}{2} & \frac{1-k}{2} & 0 & 0 \end{pmatrix} \qquad [9]$$

Thus, the obtained eigenvalues are 0, and the stability of this point cannot be determined. However, we observed that in the linearized system, and near this point, the derivative of A remains negative, indicating that the system is likely to be unstable. By examining the eigenvalues of the second set of fixed points, we found that two eigenvalues are 0, one is 1, and another is -1-Ck/4. Therefore, we concluded that there is one positive eigenvalue and one negative eigenvalue, indicating that the point is a saddle point. However, we calculated the eigenvector corresponding to the positive eigenvalue and we obtained:

$$(C(\tfrac{1-k}{16} + k4, -(1-C)(\tfrac{1-k}{16}), -(2+\tfrac{Ck}{4}), 1) \qquad [10]$$

In the context of our problem, this eigenvector has a negative third coordinate and a positive first coordinate. Since these vectors correspond to fixed points with A=I=0, we observed that the repulsive direction is in a dimension with one of the values A or I being negative, which is not feasible in the context of our problem. Therefore, we will only observe attractive behaviors.

In summary, we found an unstable fixed point where all agents are apathetic (A=1, C=I=W=0) and a subspace of fixed point given by W=A=0. The stability analysis of this subset exhibits saddle-like behavior. However, in the natural domain of the variables (all four quantities are constrained to proportions ranging from 0 to 1), these points become stable.

We also developed and solved the equations for the non-repulsive variant of the model. The terms involving repulsive interaction are not present and the dynamic equations are as follows:

$$\frac{dC}{dt} = WC\frac{k}{4} \qquad [11]$$

$$\frac{dI}{dt} = 0 \qquad [12]$$

$$\frac{dA}{dt} = 0 \qquad [13]$$

$$\frac{dW}{dt} = -WC\frac{k}{4} \qquad [14]$$

In this variant, incoherent and apathetic communities, on average, do not change over time. When k>0, all weak agents become coherent, and, as a consequence, their communities' final proportions of agents do not depend on k. Fig. 2B shows the numerical solutions for the non-repulsive model. Additionally, we performed computational simulations for both variants of the model and obtained the same results (see appendix for details).

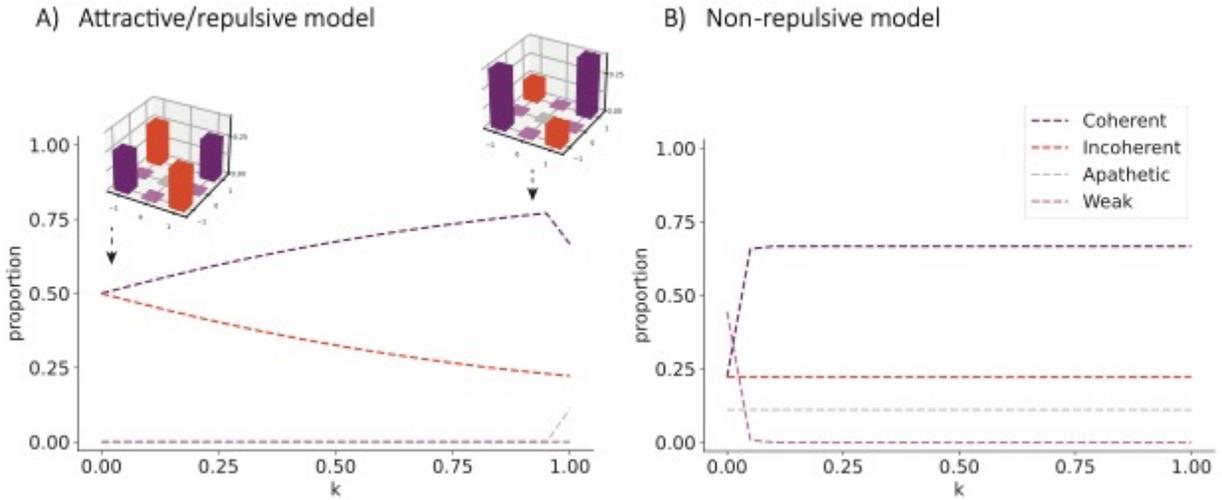

**Figure 2: Final states.** The model's final states are shown for different values of k. **A)** Attractive-repulsive model: The figure depicts the numerical solution of the model. The coherent community is shown in purple, the incoherent one in orange, the apathetic community in gray, and the weak one in pink. For k<1, as k increases, so does the final proportion of coherent agents. The top-left inset shows the agents' opinions' mean final distribution for k=0 and the top-right inset for k=0.9. **B)** Non-repulsive model: The figure depicts the numerical solution of the non-repulsive model. The coherent community is shown in purple, the incoherent one in orange, the apathetic community in gray, and the weak one in pink. For k>0, the final proportions of coherent agents do not vary significantly with k.

**COMPARISON WITH EMPIRICAL DATA**

Here, we focus on one of the multiple phenomena of political polarization: ideological sorting, in which opinions on seemingly unrelated topics display a strong correlation with each other. In the context of the proposed model, higher ideological sorting corresponds to an increase in the proportion of coherent agents. Next, we analyze the extent to which actual opinions on a wide variety of controversial issues are sorted and can be analyzed in the context of the developed model to study the role of homophily and pairwise coherence in opinion dynamics. We work with multiple data sources with 24,035 responses on 67 different polarizing topics. All responses indicate whether participants agree or not to each different issue (Table 1).

**Data sources**
**Zimmerman et al. 2022**. They conducted multiple behavioral experiments in order to understand how people perceive politically coherent individuals. First, they performed an online survey asking 180 participants from Argentina for their opinion on 28 different topics to select relevant and controversial political issues. This survey allowed them to select the five political issues and the five non-political topics they used in an experiment with 5,038 participants.
**ANES 2020 and 2016**. The American National Election Studies (ANES) are nationally representative surveys of eligible American voters. Surveys have been conducted before and after every presidential election since 1948. All the questions are related to US politics. We only used the survey's questions that express participants' approval or disapproval of controversial issues, and we worked with data from the last two surveys. The survey performed before the 2020 presidential election was completed by 8,280 citizens, while the one before the 2016 election by 4,270.
**Freira et al. 2021**. They performed an online behavioral experiment in 2020 in four different countries to understand how partisan differences influenced the COVID-19 pandemic perception. 1,995 participants from Argentina, Brazil, Uruguay, and the USA expressed their opinion on eight different pandemic preventive political policies.
**Pew Research 2020 and 2014 surveys**. The Pew Research Center is an American think tank that provides information on social issues and public opinion trends. Their different representative surveys cover a wide variety of topics, such as U.S. politics, climate, religion, and driverless vehicles. For our purposes, we selected two surveys: one oriented to understand the public's opinion on American federal agencies and the other to study how religion influences the daily lives of Americans. The first one was performed in 2020 and 1,013 participants completed an omnibus survey expressing whether they favor or not different American federal organizations. In the 2014 trends panel survey, 3,278 Americans completed a self-administered web survey covering a wide range of topics including religion and personal opinions. Within this survey, we focused on personal opinions on topics such as lying or meditation.

| Source and year | Number of selected issues | Number of participants | Country | Topic |
|---|---|---|---|---|
| **Zimmerman et al. 2022** | 21 | 5,218 | Argentina | Political & non-political |
| **ANES 2020** | 14 | 8,280 | USA | Political |
| **ANES 2016** | 28 | 4,270 | USA | Political |
| **Freira et al. 2021** | 8 | 1,976 | Argentina, Brazil, Uruguay & USA | Political |
| **Pew Research 2020** | 5 | 1,013 | USA | Political |
| **Pew Research 2014** | 5 | 3,278 | USA | Non-Political |

**Table 1**: A summary of the most relevant characteristics of the six data sources considered in this work: the source, the year in which the survey was conducted, the number of selected issues, the number of participants who completed each survey, the country, and whether topics were considered related to politics or not. These data sources include 24,035 responses on 67 different topics.

Although every popular opinion is of interest to political science, we observed that some of the selected datasets cover topics that are part of specific partisan agendas, e.g., abortion or gun control, while others do not relate to any political platform, e.g., pets or food preferences. To confirm that our datasets included both political and non-political issues and to determine whether the distinction between these two categories is not arbitrary, we conducted a preregistered online survey. In this survey, participants on Prolific were asked to what extent they consider the statements from each dataset to be related to politics on a 7-point Likert scale ranging from "not at all related" to "extremely related". Each participant (N=100 US citizens, 50 male/female, mean age: 38.8, s.d.: 12.6) was presented with five statements randomly chosen from six different data sources: two datasets from Zimmerman et al. 2022, ANES 2020 or ANES 2016, Freira et al. 2021, Pew Research 2020, and Pew Research 2014 (Table 1). Participants were instructed to evaluate the set of statements as a whole rather than the individual statements within it. The study received approval from the ethics committee for scientific and technological research at the Universidad Abierta Interamericana (protocol number 0-1104), and informed consent was obtained from all participants. Consistent with the preregistration, participants rated the datasets we had considered to be related to politics higher (M=5.66±0.08) compared to the non-political ones (M=1.30±0.06; linear mixed-effects model: b=4.4±0.1, t(499)=39.7, p=$2\times10^{-156}$). This result supports a consensus-based distinction between political and non-political issues (see appendix for details).

**Sorting**

For each dataset, we selected the questions that express participants' opinions on a particular topic. Moreover, we only considered controversial issues in which there is not a majoritarian opinion. The selected issues' responses exhibit high variance following our selection criteria ($\sigma^2$>0.5). Following the proposed model, all responses were mapped into three possible answers: -1, 0, or 1, where 0 expresses an undecided or neutral posture. For some questions this procedure was trivial, but for others, we merged different degrees of acceptance into one unique alternative. For example, all the following answers were considered as an agreement in our model: 'agree strongly', 'agree somewhat', 'very favorable', and 'mostly favorable'. This approach was applied to the responses from the ANES data. However, to ensure that our results were not influenced by this classification method, we created two datasets for each ANES survey (2016 and 2020): one following the described merging procedure and another dataset that includes only answers to questions where participants had only three possible responses, thus eliminating the need to collapse data. This approach allows us to control for the influence of response categorization on our findings (see appendix for details). Furthermore, in order to observe ideological alignment, we needed to code responses as in the model: right-wing opinions as 1 and left-wing opinions as -1 regardless of how the question was presented. To do so, we followed the procedure proposed by Zimmerman et. al 2022 where they projected the agree/disagree opinions to the data's first principal component, which, in the political domain, is equivalent to coding them as left/right-wing answers and align the responses into their corresponding ideology.

Once we had responses on controversial issues categorized into three possible opinions (typical left-wing, neutral, and typical right-wing), we proceeded to contrast the data's partisan-ideological sorting with the proposed two-dimensional model. This analysis focused on the relationship between the proportion of coherent (C) and incoherent (I) opinions. For each dataset, we examined every possible pair of opinions to identify the number of participants presenting coherent opinions (both opinions aligning as 1,1 or -1,-1) as opposed to incoherent opinions (1,-1 or -1,1). For each dataset, we calculated a single sorting value for each pair of opinions, which is the proportion of participants expressing coherent opinions relative to the total number of both coherent and incoherent opinions. Then, we computed the average sorting value for each dataset by considering all possible topic pair combinations. Therefore, sorting (S) was determined by averaging the proportion of pairwise coherent opinions in relation to the combined total of coherent and incoherent opinions across all potential opinion pairs within each dataset (see appendix for details).

$$S = \frac{C}{(C+I)} \quad [15]$$

Fig. 3A shows the mean sorting value for each dataset ranked from lowest to highest. Non-political datasets are shown in light gray and political datasets in black. Noteworthy, the political datasets exhibit the highest sorting values.

Furthermore, we followed the same procedure to compute the mean sorting value of 100 simulations where k=0, these are homophilic-only simulations. First, for each simulation's final state, we projected the opinions to their first principal component, and then we computed the corresponding sorting value. The mean sorting value of the simulations (y=0.52) is depicted in a dashed horizontal gray line (Fig. 3A). We observed that none of the non-political sorting values are significantly different from homophilic-only simulations (one sample t-test: ts<3.1, ps>0.013) and 10 out of 12 political sorting values are (ts>5.7, ps<0.0004).

Additionally, having the model's numerical solution allowed us to obtain the model's k value that results in a final state matching the sorting value observed in each dataset. Because the proportion of the coherent community, and therefore sorting, increases monotonically with respect to k for k<1, we were able to map each sorting value to a corresponding k-value. For each dataset, we used the range provided by the standard error of the mean sorting to determine the minimum and maximum k-values, and then we computed the mean k-value. This parameter allows us to quantify an underlying social mechanism derived from observed opinion data. The value of k quantifies the relevance of pairwise-coherence favoritism in relation to homophily across various contexts, such as different issues, countries, or populations. A k-value of 0 would indicate that the observed data could be explained solely by considering homophily. A value of 0.5 would highlight that pairwise-coherence favoritism is as relevant as homophilic interactions. A k-value of 1 does not take into account the influence of homophily, which contradicts well-established findings in social science. Therefore, we focused our analysis on cases where k<1. For each dataset, we derived a k-value from the model that could explain the observed levels of sorting. Moreover, through the online survey, we obtained their perceived politicalness. The results of social experiments have shown that pairwise-coherence favoritism is present in political discussions but not in non-political issues (36). Consistent with this finding, we expected higher k-values among datasets categorized as political. As preregistered, we found that the datasets' k-values were correlated with their mean perceived politicalness reported in the online survey (Spearman correlation: r=0.62, p=0.03; Fig. 3B).

Taken altogether, by combining the model with actual data, we found that non-political controversial opinions exhibit the lowest observed levels of sorting, which can be explained by homophily and attractive-repulsive interactions. But, in the political domain, homophily alone cannot explain the emergence of the observed levels of ideological sorting. These results are in accordance with recent experimental work that showed that both homophily and pairwise-coherence favoritism are relevant in pair-wise political interactions. Moreover, our model allowed us to study how these mechanisms impact the macroscopic opinion landscape. Particularly, how they relate to issue-polarization and ideological alignment.

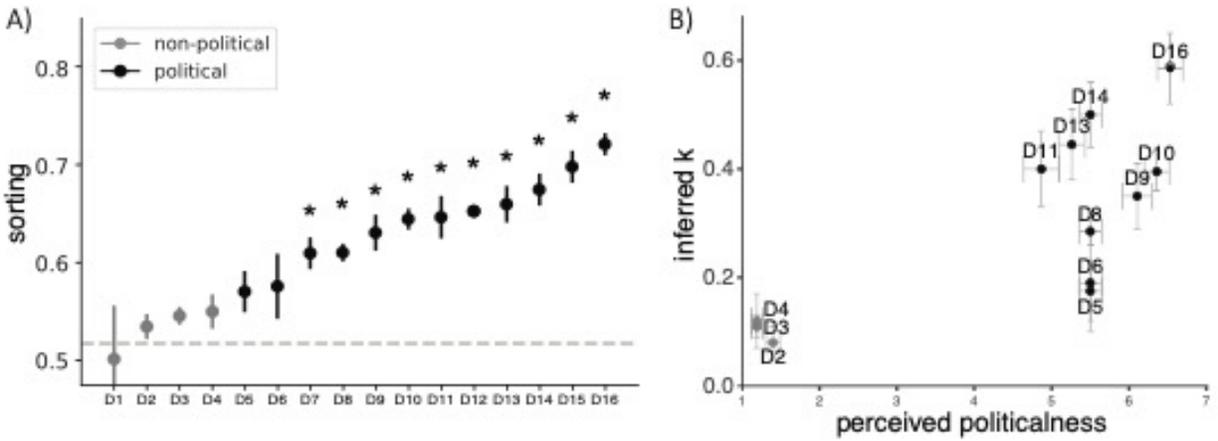

**Figure 3: Sorting patterns, model's parameter k, and perceived politicalness. A)** The figure shows the mean sorting value and its standard error (SEM) for each dataset ranked from lowest to highest. Dataset ID numbers correspond to different sets of questions, as detailed in Table S1. Non-political datasets are shown in light gray and political datasets in black. Asterisks indicate whether the sorting value is significantly higher than the one observed in homophilic-only simulations (*p<0.01). **B)** The figure shows the perceived politicalness obtained from the online survey (x-axis), and the model's parameter k obtained from the opinions in each dataset through the model (y-axis). The figure shows the mean and SEM of these variables for each dataset. Non-political datasets are shown in light gray and political datasets in black.

## DISCUSSION

In this work, we considered a multidimensional agent-based model to study how pairwise-coherence favoritism, a novel finding from experimental psychology, could explain the emergence of ideological sorting. This model considers two well-known and studied phenomena: agents who share similar opinions are more likely to interact and these pairwise interactions can be attractive or repulsive depending on their similarity. Moreover, we included a third and novel finding in which agents are more attracted to coherent ingroups rather than incoherent or outgroup members. Interestingly, by incorporating this last assumption, opinions become more aligned, and we were able to reproduce different correlation patterns similar to those widely observed among political opinions, i.e., ideological sorting.

We developed a two-dimensional model as our main interest was to study the effect of pairwise-coherence favoritism in political interactions and its impact on ideological sorting. The commonly used one-dimensional models of opinion formation lack the possibility to model these mechanisms as well as other relevant phenomena involving the relation between topics. The implementation of opinion models necessarily requires making assumptions to decide which dynamics should be implemented and how. Here, the criterion was to consider only mechanisms that are experimental findings from social psychology and not merely hypothetical assumptions. Additionally, we compared the model's outcomes with 24,035 opinions on controversial issues. These opinions belong to multiple surveys that studied a wide diversity of topics in different countries. As we formulated and numerically integrated the model's master equations, we were able to obtain the model's parameter k. This allowed us to replicate the opinions' distribution for every dataset, observe that there are different mechanisms involved in political and non-political discussions, and study which underlying psychological mechanisms could be driving the observed levels of ideological sorting. For instance, for the non-political datasets, k was not found to be significantly different from 0, implying that pairwise coherence does not influence these debates. These results are consistent with previous studies that showed that pairwise coherence is characteristic of the political domain.

Moreover, our analysis helped us understand the relationship between the different data sources and the traditional political parties' agendas. For example, the ANES' results, which cover the most relevant and discussed topics within American politics, exhibit a higher proportion of coherent respondents. On the other end, the least aligned political topics are related to the COVID-19 pandemic perception (58). On the one hand, opinions regarding the pandemic are recent and they are not expected, a priori, to be aligned with a particular political platform. On the other hand, the required response to control the virus' spread is not merely political, as it involves both political and personal aspects of every citizen. Also, we know that the perception and compliance with the recommended behavior varied widely across different countries and contexts (59–62).

The simplicity of the proposed model comes with its own limitations. Firstly, we focused our study on controversial issues that exhibit a low proportion of undecided opinions. Whether and how the proposed model could explain other opinions' distributions remains unexplored. Secondly, the model assumes that each agent holds an opinion on two topics. Extending the model to more topics has yet to be explored and presents unique challenges. One challenge is defining the incoherent population when the number of topics is odd, as it becomes impossible to have an equal number of views from each political leaning (positive and negative opinions). Additionally, when exploring a higher number of topics, there will always be two states of coherent agents (all positive or all negative opinions). In contrast, the incoherent ones can be represented by multiple states, each having an equal number of positive and negative opinions. This complexity should be taken into account when calculating sorting values. Thirdly, although it allows us to capture the temporal dynamics of opinions, it is hard to define an adequate time frame in order to match the model's time evolution with actual observed opinion changes. Finally, we observed that it is not possible to replicate the correlation patterns of political opinions by only considering attractive and repulsive homophilic interactions. But this observation does not necessarily imply that pairwise-coherence favoritism is the one and only mechanism driving ideological alignment. More behavioral experiments are needed to understand the importance of each of the multiple factors associated with ideological sorting.

It has become urgent to find and implement solutions to the worryingly rising political polarization. As issue polarization, outgroup hate and distrust, and political segregation increase, democracy gets weaker. One key element of this herculean mission is to fully understand the underlying social, political, and psychological mechanisms driving polarization. To do so, more combined efforts from experimental approaches, theoretical modeling, and large-scale empirical studies are needed.

## DATA AVAILABILITY

All simulation data and code for simulations, figures, and the numeric solutions are available at GitHub
(https://github.com/lupipedraza/Attraction-by-ingroup-coherence-explains-the-emergence-of-ideological-sorting).
This work analyzes data from the following papers and projects: Zimmerman, F., Garbulsky, G., Ariely, D., Sigman, M. & Navajas, J. Political coherence and certainty as drivers of interpersonal liking over and above similarity. Sci. Adv. 8, eabk1909 (2022). Freira, L., Sartorio, M., Boruchowicz, C., Lopez Boo, F. & Navajas, J. The interplay between partisanship, forecasted COVID-19 deaths, and support for preventive policies. Humanit. Soc. Sci. Commun. 8, 1-10 (2021). American National Election Studies. 2021. ANES 2020 Time Series Study Full Release [dataset and documentation]. February 10, 2022 version. www.electionstudies.org
https://www.pewresearch.org/politics/dataset/march-2020/
https://www.pewresearch.org/religion/dataset/american-trends-panel-wave-6/

This manuscript was posted on a preprint: arXiv:2304.12559.

## FUNDING

This research was supported by the James McDonnell Foundation 21st Century Science Initiative in Understanding Human Cognition—Scholar Award (Grant No. 220020334) and by the Templeton World Charity Foundation (Grant No. TWCF-2022-31322).

## APPENDIX

**Agents' possible opinions**

We investigated additional variations of the model, specifically a scenario in which an agent can hold five different opinions per topic: strongly in favor (+2), somewhat in favor (+1), indifferent (0), somewhat against (-1), and strongly against (-2). Given that each agent holds an opinion on two topics, this results in each agent having 25 possible states (Fig. S1A). We considered agents as coherent if their opinions are either both positive or both negative. Conversely, agents are considered incoherent if they hold one positive and one negative opinion. To determine the final states of the model, we conducted simulations varying k in steps of 0.1. Fig. S1B shows the final proportions of coherent and incoherent agents as a function of k. The behavior of this model is consistent with that observed in the simpler three-opinion model: at k=0, the proportions of coherent and incoherent agents are equal, and, while k<1, as k increases, the proportion of coherent agents rises.

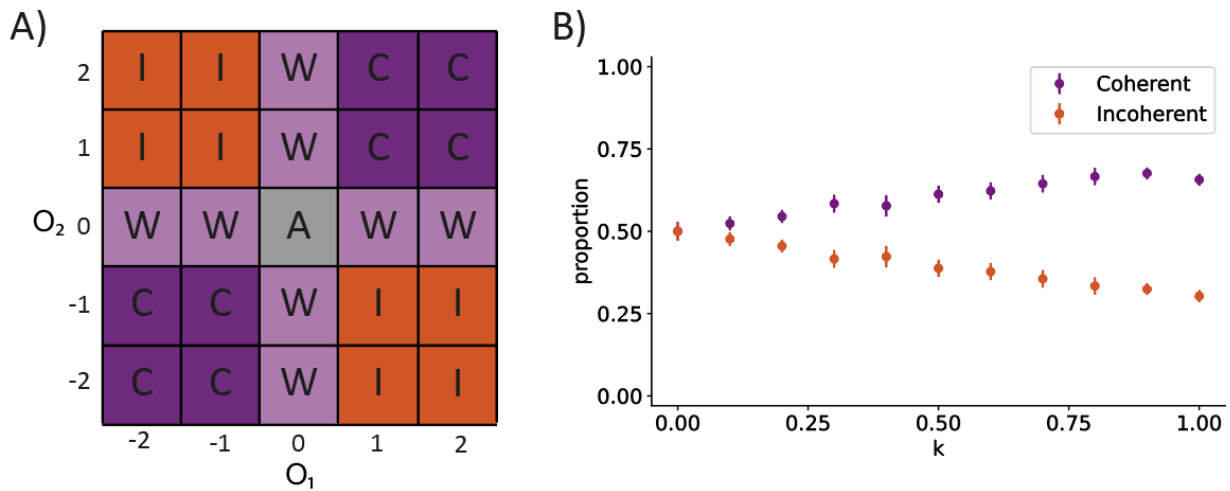

**Figure S1: Five-opinion model**. **A)** Each agent holds an independent opinion on two different topics and according to their opinions they are classified in four different communities: coherent (C), incoherent (I), weak (W), or apathetic (A). **B)** The figure depicts the mean (dots) and the standard deviation (error bar) values of the final populations' proportions in the five-opinion model. The coherent community is shown in purple and the incoherent one in orange.

**Attraction-similarity boundary**

We explored further variations of the model, focusing on the attraction-similarity threshold (T). This threshold determines the degree of similarity at which agents either attract or repel each other. Agents are attracted to each other if their similarity is equal to or exceeds T; otherwise, they repel each other. Given that agents express bidimensional discrete opinions, the distance between any two agents can only take values from the set {0, ¼, ½, ¾, 1}. We did not explore the extreme cases 0 and 1 where agents either always attract or only identical opinions attract.

Therefore, we focused on intermediate thresholds of ¼, ½, and ¾. Fig. S2A illustrates these thresholds, showing for each scenario the opinions that attract (shown in orange) and repel (gray) each other for an agent holding two negative opinions (purple dot).

We derived the corresponding equations for each threshold and performed numerical integration to explore their dynamics. Fig. S2B shows the final population distributions for different values of k. Notably, at lower values of T (½ and ¼), the final states are concentrated either in the apathetic or in the coherent populations, a scenario that is not observed in actual opinion data.

Moreover, to test the robustness of our findings, we considered an alternative metric for defining similarity. Instead of the Manhattan distance, we considered the L-infinity distance, which measures the maximum absolute difference between the components of two vectors. We set the threshold at ½, creating the conditions illustrated in Fig. S2A. We observed that in the final state, for k<1, all communities vanish, and there are only coherent and incoherent individuals (Fig. S2B). In this scenario, all points where there are coherent and incoherent individuals represent fixed points according to our model equations.

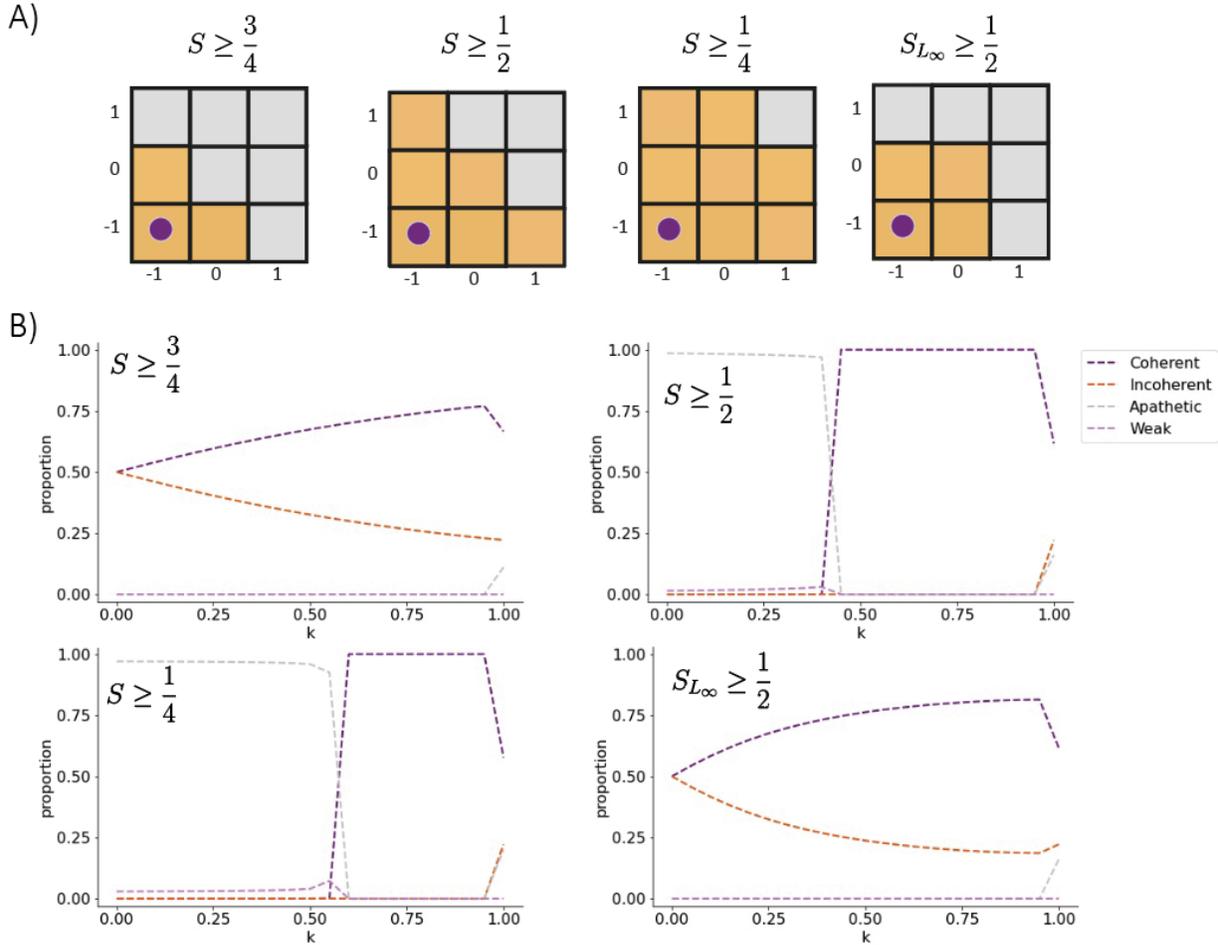

**Figure S2**: We explored different values for the attraction-similarity boundary, T. **A)** The figure shows the states that are attractive (in yellow) and those that generate rejection (in gray) for an agent with opinion (-1, -1), as marked in purple, for each value of T. **B)** The figure depicts the numerical solution for each model variant. It shows the final proportion of coherent agents in purple, the final proportion of incoherent agents in orange, the final proportion of apathetic agents in gray, and the final proportion of weak agents in pink.

**Simulations**

We present the results of the model's simulations. All simulations were done for systems with N=1000 agents and we ran 100 simulations per set of parameters. The parameter k was varied from 0 to 1 by steps of 0.05. Simulations stopped when the system reached the stationary state, and no agent could move from one community to another. We observed the proportion of agents in each community over time and focused on the system's final state. For every k<1, weak and apathetic communities disappear. Fig. S3 shows the final distribution of coherent and incoherent communities for different k values, over the results obtained from the master equations. This visualization clearly illustrates the correspondence between the two sets of results. When k=0 the proportion of coherent and incoherent agents are the same ($C(t_f)=I(t_f)=0.5$) and there is no correlation between opinions. As k increases, so does the proportion of coherent agents. In the specific case where interactions are solely driven by pairwise coherence (k=1), apathetic and incoherent agents do not interact with any other population, so these groups do not change over time.

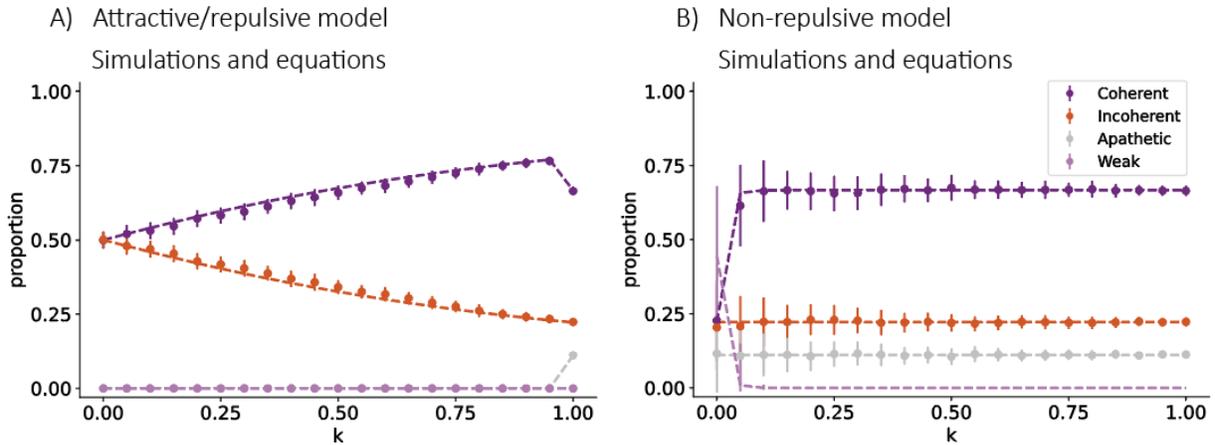

**Figure S3**: **Final states.** The model's final states are shown for different values of k. The figure depicts the mean (dots) and the standard deviation (error bars) values of the final populations' proportions in the simulations and the numeric solution of the equations (lines). The coherent community is shown in purple, the incoherent one in orange, the apathetic community in gray, and the weak one in pink. **A)** The final states of the attractive-repulsive model. **B)** The final states of the non-repulsive model.

**Online survey**

The online survey had two goals. First, it aimed to determine if there is a unified criterion for identifying whether various issues are considered "political." Secondly, the survey attempted to examine whether the extent to which issues are perceived as political correlates with a model-based measure of ideological sorting.

Participants: The study received approval from the ethics committee for scientific and technological research at the Universidad Abierta Interamericana prior to data collection (protocol number 0-1104). As preregistered (https://aspredicted.org/wp2df.pdf), we targeted 100 participants, balanced in terms of gender, recruited through Prolific. We anticipated capturing a medium-to-large effect size. Thus, a sample of 100 participants would achieve a power of over 0.90 with an alpha level of 0.05. Our sample included 100 U.S. citizens (50 male and 50 female, mean age: 38.8 years, s.d.: 12.6 years). Each participant provided informed consent and correctly answered two attention checks throughout the study.

Stimuli: We considered all groups of statements from the studies included in this work (see Table 1). Notably, Zimmerman et al. 2022 conducted multiple surveys, providing us with three distinct groups of statements: two related to politics and one unrelated to politics. This resulted in a total of eight groups of statements. The number of statements within each group ranged from 5 to 28. To unify the presented stimuli, we randomly selected five statements from each group. Additionally, some datasets contained very similar statements; in such cases, we presented only one of them, randomly selected. This selection process was applied to both the political online surveys by Zimmerman et al. 2022 and the ANES surveys from 2016 and 2020. So, for example, each participant saw five statements from either the ANES 2016 or the ANES 2020. In summary, each participant was presented with six datasets, each containing five statements.

Measures: For each dataset, participants were asked to evaluate each set as a whole. They were asked: "To what extent do you consider these statements to be related to politics?" Responses were collected on a 7-point Likert scale, ranging from "Not at all related" to "Extremely related".

Results: As preregistered, we conducted two analyses. First, we performed a linear mixed-effects model analysis with the dependent variable being the perceived relation to politics. The model included a fixed effect for a dummy variable indicating whether the dataset has been considered related to politics, and participants' IDs as a random effect. Participants rated the datasets that had been considered to be related to politics higher (M=5.66±0.08) compared to the non-political ones (M=1.30±0.06; linear mixed-effects model: b=4.4±0.1, t(499)=39.7, $p=2\times10^{-156}$). Secondly, we performed a Spearman correlation analysis between the mean ratings per dataset and the mean k-values obtained from the model, revealing a significant correlation (Spearman correlation: r=0.62, p=0.03; see section 'Datasets for sorting' in the appendix for details).

**Datasets for sorting**

Here we explain how we obtained the 16 datasets shown in Fig. 3A from the six studies listed in Table 1. For each ANES survey (2016 and 2020), we created two datasets to ensure that our results were not influenced by the classification method. One dataset included all questions that capture participants' opinions on specific topics (all opinion questions, see Table S1), while the other dataset includes only those questions offering three response options (ternary choice questions). Moreover, because sorting values are calculated based on the proportion of participants expressing coherent opinions across pairs of topics, it was crucial to ensure that these datasets included responses from the same participants on the same issues. From Freira et al. 2021, we obtained four datasets, corresponding to surveys conducted with participants in Argentina, Brazil, Uruguay, and the USA. Furthermore, Zimmerman et al. 2022 conducted four online surveys on political and non-political topics (online surveys, political and non-political), which allowed them to select the five political issues and the five non-political topics they used in an experiment (live experiment, political and non-political). Table S1 displays the most relevant characteristics of these 16 datasets.

| Number | Source and year | Set of questions | N. of issues | N. of part. | Country | Topic |
|---|---|---|---|---|---|---|
| D1 | **Zimmerman et al. 2022** | Online survey. Non-political issues A. | 4 | 88 | Argentina | Non-Political |
| D2 | **Pew Research 2014** | Non-political issues | 5 | 3,278 | USA | Non-Political |
| D3 | **Zimmerman et al. 2022** | Live experiment. Non-political issues. | 5 | 2,406 | Argentina | Non-Political |
| D4 | **Zimmerman et al. 2022** | Online survey. Non-political issues B. | 5 | 92 | Argentina | Non-Political |
| D5 | **Freira et al. 2021** | Uruguayan survey | 8 | 371 | Uruguay | Political |
| D6 | **Freira et al. 2021** | American survey | 7 | 614 | USA | Political |
| D7 | **Zimmerman et al. 2022** | Live experiment. Political issues. | 5 | 2,632 | Argentina | Political |
| D8 | **Freira et al. 2021** | Argentinian survey | 8 | 639 | Argentina | Political |
| D9 | **Zimmerman et al. 2022** | Online survey. Political issues A. | 6 | 88 | Argentina | Political |
| D10 | **ANES 2016** | Ternary choice questions | 11 | 4,270 | USA | Political |
| D11 | **Zimmerman et al. 2022** | Online survey. Political issues B. | 6 | 92 | Argentina | Political |
| D12 | **ANES 2016** | All opinion questions | 28 | 4,270 | USA | Political |
| D13 | **Pew Research 2020** | All opinion questions | 5 | 1,013 | USA | Political |
| D14 | **Freira et al. 2021** | Brazilian survey | 8 | 352 | Brazil | Political |
| D15 | **ANES 2020** | Ternary choice questions | 5 | 8,280 | USA | Political |
| D16 | **ANES 2020** | All opinion questions | 14 | 8,280 | USA | Political |

**Table S1**: The most relevant characteristics of all the datasets used in this work: the source, the set of questions, the year in which the survey was conducted, the number of selected issues, the number of participants who completed each survey, the country, and whether the topics were considered related to politics or not.

**Overview of all the analyzed questions**

**Zimmerman et al. 2022 - Live experiment. Political issues**

1. High school students should be allowed to go on strike.
2. The government should mandate a transgender hiring quota for public servants.
3. The government should subsidize the broadcasting of Argentine football matches.
4. National universities should start charging a fee to those who can afford it.
5. Argentina should sign bilateral trade agreements with the United States.

**Zimmerman et al. 2022- Online survey. Political issues A**

1. In order to progress, Argentina should imitate developed countries.
2. The government should mandate a transgender hiring quota for public servants.
3. No union claim should interfere with the free movement of persons.
4. The government should subsidize the broadcasting of Argentine football matches.
5. The minimum age of criminal responsibility in Argentina (now 16 years) should be reduced.
6. The sentence for rapists should be the death penalty.

Not considered because they were not found to be controversial:
- The state should not subsidize any religious institution.
- The consumption or use of recreational drugs should not be a crime.

**Zimmerman et al. 2022 - Online survey. Political issues B**

1. If someone does graffiti on a public building, they should be detained.
2. Argentina should sign bilateral trade agreements with the United States.
3. National universities should start charging a fee to those who can afford it.
4. High school students should be allowed to go on strike.
5. People with criminal records should not receive any state subsidy.
6. Foreigners should pay for medical care in an Argentine public hospital.

Not considered because they were not found to be controversial:
- Heterosexual couples should have higher priority than gay couples to adopt.
- Abortion should not be a crime.

**Zimmerman et al. 2022 - Live experiment. Non-political issues**

1. One should choose to adopt a dog over a cat.
2. Everyone should use a bidet if they have the possibility.
3. All good barbecues must have blood sausages.
4. Baked schnitzels taste better than fried schnitzels.
5. It is better to go on holidays to the mountains rather than to the seaside.

**Zimmerman et al. 2022 - Online survey. Non-political issues A**

1. Everyone should use a bidet if they have the possibility.
2. When ordering a dozen 'medialunas' more than half should be of animal grease instead of butter.
3. In order to improve the taste, 'empanadas' should carry raisins.
4. One should choose to adopt a dog over a cat.

Not considered because they were not found to be controversial:
- Argentinian Rock radio stations should play more Redondos' songs than Soda Stereo's.
- Whenever possible one should choose the aisle seat over the window seat.

**Zimmerman et al. 2022 - Online survey. Non-political issues B**

1. All good barbecues must have blood sausages.
2. Baked schnitzels taste better than fried schnitzels.
3. It is better to go on holidays to the mountains rather than to the seaside.
4. Maradona should receive the award as 'best history Argentinian soccer player' and Messi should not.
5. 'Queso y dulce' dessert should have sweet potato and not quince.

Not considered because they were not found to be controversial:
- 'Mate' should be taken with sugar and not bitter.

**ANES 2020 - All opinion questions**

1. V201252 MEDICAL INSURANCE
Where would you place yourself on this scale, or haven't you thought much about this? From Government insurance plan to Private insurance plan

2. V201258 GOV ASSISTANCE TO BLACKS
Where would you place yourself on this scale, or haven't you thought much about this? From Government should help blacks to Blacks should help themselves

3. V201306 FEDERAL BUDGET SPENDING: TIGHTENING BORDER SECURITY
What about tightening border security to prevent illegal immigration? Should federal spending on tightening border security to prevent illegal immigration be increased, decreased, or kept the same?

4. V201309 FEDERAL BUDGET SPENDING: DEALING WITH CRIME
What about dealing with crime? Should federal spending on dealing with crime be increased, decreased, or kept the same

5. V201318 FEDERAL BUDGET SPENDING: AID TO THE POOR
What about aid to the poor? Should federal spending on aid to the poor be increased, decreased, or kept the same?

6. V201336 STD ABORTION
There has been some discussion about abortion during recent years. Which one of the opinions on this page best agrees with your view?
a) By law, abortion should never be permitted
b) The law should permit abortion only in case of rape, incest, or when the woman's life is in danger
c) The law should permit abortion other than for rape/incest/danger to woman but only after need clearly established
d) By law, a woman should always be able to obtain an abortion as a matter of personal choice

7. V201406 SERVICES TO SAME SEX COUPLES
Do you think business owners who provide wedding-related services should be allowed to refuse services to same-sex couples if same-sex marriage violates their religious beliefs, or do you think business owners should be required to provide services regardless of a couple's sexual orientation?

8. V201409 TRANSGENDER POLICY
Should transgender people - that is, people who identify themselves as the sex or gender different from the one they were born as - have to use the bathrooms of the gender they were born as, or should they be allowed to use the bathrooms of their identified gender?

9. V201415 GAY AND LESBIAN COUPLES BE ALLOWED TO ADOPT
Do you think gay or lesbian couples should be legally permitted to adopt children?

10. V201416 POSITION ON GAY MARRIAGE
Which comes closest to your view?
a) Gay and lesbian couples should be allowed to legally marry
b) Gay and lesbian couples should be allowed to form civil unions but not legally marry
c) There should be no legal recognition of gay or lesbian couples' relationship

11. V201417 US GOVERNMENT POLICY TOWARD UNAUTHORIZED IMMIGRANTS
Which comes closest to your view about what government policy should be toward unauthorized immigrants now living in the United States?
a) Make all unauthorized immigrants felons and send them back to their home country
b) Have a guest worker program that allows unauthorized immigrants to remain in US to work but only for limited time
c) Allow unauthorized immigrants to remain in US & eventually qualify for citizenship but only if they meet requirements
d) Allow unauthorized immigrants to remain in US & eventually qualify for citizenship without penalties

12. V201418 FAVOR OR OPPOSE ENDING BIRTHRIGHT CITIZENSHIP
Some people have proposed that the U.S. Constitution should be changed so that the children of unauthorized immigrants do not automatically get citizenship if they are born in this country. Do you favor, oppose, or neither favor nor oppose this proposal?

13. V201424 FAVOR OR OPPOSE BUILDING A WALL ON BORDER WITH MEXICO
Do you favor, oppose, or neither favor nor oppose building a wall on the U.S. border with Mexico?

14. V201429 BEST WAY TO DEAL WITH URBAN UNREST
What is the best way to deal with the problem of urban unrest and rioting? Some say it is more important to use all available force to maintain law and order, no matter what results. Others say it is more important to correct the problems of racism and police violence that give rise to the disturbances. And, of course, other people have opinions in between. From Solve problems of racism and police violence to Use all available force to maintain law and order

Not considered because they were not found to be controversial:

- V201300 FEDERAL BUDGET SPENDING: SOCIAL SECURITY
What about Social Security? Should federal spending on Social Security be increased, decreased, or kept the same?

- V201401 GOVERNMENT ACTION ABOUT RISING TEMPERATURES
Do you think the federal government should be doing more about rising temperatures, should be doing less, or is it currently doing the right amount?

- V201412 LAWS PROTECT GAYS/LESBIANS AGAINST JOB DISCRIMINATION
Do you favor or oppose laws to protect gays and lesbians against job discrimination?

**ANES 2020 - Ternary choice questions**

1. V201406 SERVICES TO SAME SEX COUPLES
Do you think business owners who provide wedding-related services should be allowed to refuse services to same-sex couples if same-sex marriage violates their religious beliefs, or do you think business owners should be required to provide services regardless of a couple's sexual orientation?

2. V201409 TRANSGENDER POLICY
Should transgender people - that is, people who identify themselves as the sex or gender different from the one they were born as - have to use the bathrooms of the gender they were born as, or should they be allowed to use the bathrooms of their identified gender?

3. V201424 FAVOR OR OPPOSE BUILDING A WALL ON BORDER WITH MEXICO
Do you favor, oppose, or neither favor nor oppose building a wall on the U.S. border with Mexico?

4. V201418 FAVOR OR OPPOSE ENDING BIRTHRIGHT CITIZENSHIP
Some people have proposed that the U.S. Constitution should be changed so that the children of unauthorized immigrants do not automatically get citizenship if they are born in this country. Do you favor, oppose, or neither favor nor oppose this proposal?

5. V201415 GAY AND LESBIAN COUPLES BE ALLOWED TO ADOPT
Do you think gay or lesbian couples should be legally permitted to adopt children?

**ANES 2016 - All opinion questions**

1. V161113 Healthcare
Do you favor, oppose, or neither favor nor oppose the health care reform law passed in 2010? This law requires all Americans to buy health insurance and requires health insurance companies to accept everyone.

2. V161184 Insurance
Where would you place yourself on this scale, or haven't you thought much about this? From Government insurance plan to Private insurance plan.

3. V161196 Wall
Do you favor, oppose, or neither favor nor oppose building a wall on the U.S. border with Mexico?

4. V161198 Black
Where would you place yourself on this scale, or haven't you thought much about this? From Government should help Blacks to Blacks should help themselves.

5. V161201 Environment
Where would you place yourself on this scale, or haven't you thought much about this? From Regulate business to protect the environment and create jobs to No regulation because it will not work and will cost jobs.

6. V161214 Syrian
Do you favor, oppose, or neither favor nor oppose allowing Syrian refugees to come to the United States?

7. V161228 TransBathroom
Should transgender people – that is, people who identify themselves as the sex or gender different from the one they were born as – have to use the bathrooms of the gender they were born as, or should they be allowed to use the bathrooms of their identified gender?

8. V161227 Same Sex Service
Do you think business owners who provide wedding-related services should be allowed to refuse services to same-sex couples if same-sex marriage violates their religious beliefs, or do you think business owners should be required to provide services regardless of a couple's sexual orientation?

9. V161193 Birthright
Some people have proposed that the U.S. Constitution should be changed so that the children of unauthorized immigrants do not automatically get citizenship if they are born in this country. Do you favor, oppose, or neither favor nor oppose this proposal?

10. V161204 Affirmative Action
Do you favor, oppose, or neither favor nor oppose allowing universities to increase the number of black students studying at their schools by considering race along with other factors when choosing students?

11. V161213 ISIS
Do you favor, oppose, or neither favor nor oppose the U.S. sending ground troops to fight Islamic militants, such as ISIS, in Iraq and Syria?

12. V161229 Gay protection
Do you favor or oppose laws to protect gays and lesbians against job discrimination?

13. V161232 Abortion
There has been some discussion about abortion during recent years. Which one of the opinions on this page best agrees with your view?
a) By law, abortion should never be permitted.
b) By law, only in case of rape, incest, or woman's life in danger.
c) By law, for reasons other than rape, incest, or woman's life in danger if need established.
d) By law, abortion as a matter of personal choice.

14. V161233 death penalty
Do you favor or oppose the death penalty for persons convicted of murder?

15. V161343 protesters
When protestors get 'roughed up' for disrupting political events, how much do they generally deserve what happens to them?

16. V161346 feminism
How well does the term feminist' describe you?

17. V162123 Countries like America
'The world would be a better place if people from other countries were more like Americans.' Do you [agree strongly, agree somewhat, neither agree nor disagree, disagree somewhat, or disagree strongly / disagree strongly, disagree somewhat, neither agree nor disagree, agree somewhat, or agree strongly] with this statement?

18. V162169 forefathers
'Our country would be great if we honor the ways of our forefathers, do what the authorities tell us to do, and get rid of the 'rotten apples' who are ruining everything.' (Do you [agree strongly, agree somewhat, neither agree nor disagree, disagree somewhat, or disagree strongly / disagree strongly, disagree somewhat, neither agree nor disagree, agree somewhat, or agree strongly] with this statement?)

19. V162170 strong leader
What our country really needs is a strong, determined leader who will crush evil and take us back to our true path.' (Do you [agree strongly, agree somewhat, neither agree nor disagree, disagree somewhat, or disagree strongly / disagree strongly, disagree somewhat, neither agree nor disagree, agree somewhat, or agree strongly] with this statement?)

20. V162210 traditional family
'This country would have many fewer problems if there were more emphasis on traditional family ties.' (Do you [agree strongly, agree somewhat, neither agree nor disagree, disagree somewhat, or disagree strongly / disagree strongly, disagree somewhat, neither agree nor disagree, agree somewhat, or agree strongly] with this statement?)

21. V162211 help blacks
'Irish, Italians, Jewish and many other minorities overcame prejudice and worked their way up. Blacks should do the same without any special favors.' Do you [agree strongly, agree somewhat, neither agree nor disagree, disagree somewhat, or disagree strongly / disagree strongly, disagree somewhat, neither agree nor disagree, agree somewhat, or agree strongly] with this statement?

22. V162221 hispanics
How important is it that more Hispanics be elected to political office?

23. V162244 equality
'This country would be better off if we worried less about how equal people are.' (Do you [agree strongly, agree somewhat, neither agree nor disagree, disagree somewhat, or disagree strongly / disagree strongly, disagree somewhat, neither agree nor disagree, agree somewhat, or agree strongly] with this statement?)

24. V162266 traditions
Now thinking about minorities in the United States. Do you [agree strongly, agree somewhat, neither agree nor disagree, disagree somewhat, or disagree strongly / disagree strongly, disagree somewhat, neither agree nor disagree, agree somewhat or agree strongly] with the following statement? 'Minorities should adapt to the customs and traditions of the United States'

25. V162268 immigrants economy
And now thinking specifically about immigrants. (Do you [agree strongly, agree somewhat, neither agree nor disagree, disagree somewhat, or disagree strongly /disagree strongly, disagree somewhat, neither agree nor disagree, agree somewhat or agree strongly] with the following statement?) 'Immigrants are generally good for America's economy.'

26. V162270 immigrants crime
(Do you [agree strongly, agree somewhat, neither agree nor disagree, disagree somewhat, or disagree strongly /disagree strongly, disagree somewhat, neither agree nor disagree, agree somewhat or agree strongly] with the following statement?) 'Immigrants increase crime rates in the United States.'

27. V162276 differences
Please say to what extend you agree or disagree with the following statement: 'The government should take measures to reduce differences in income levels'. (Do you [agree strongly, agree somewhat, neither agree nor disagree, disagree somewhat, or disagree strongly / disagree strongly, disagree somewhat, neither agree nor disagree, agree somewhat or agree strongly]?)

28. V162295 torture
Do you favor, oppose, or neither favor nor oppose the U.S. government torturing people who are suspected of being terrorists, to try to get information?

Not considered because they were not found to be controversial:

- V161154 Military
How willing should the United States be to use military force to solve international problems?

- V161226 Parental leave
Do you favor/oppose, or neither favor nor oppose requiring employers to offer paid leave to parents of new children?

- V162125x Flag
How good/bad does R feel to see American flag?

- V162150x equal pay
Favor/oppose equal pay for men and women

- V162168 free thinkers
'Our country needs free thinkers who will have the courage to defy traditional ways, even if this upsets many people.' Do you [agree strongly, agree somewhat, neither agree nor disagree, disagree somewhat, or disagree strongly / disagree strongly, disagree somewhat, neither agree nor disagree, agree somewhat, or agree strongly] with this statement?

- V162186 business regulation
How much government regulation of business is good for society?

**ANES 2016 - Ternary choice questions**

1. V161113 Healthcare
Do you favor, oppose, or neither favor nor oppose the health care reform law passed in 2010? This law requires all Americans to buy health insurance and requires health insurance companies to accept everyone.

2. V161196 Wall
Do you favor, oppose, or neither favor nor oppose building a wall on the U.S. border with Mexico?

3. V161214 Syrian
Do you favor, oppose, or neither favor nor oppose allowing Syrian refugees to come to the United States?

4. V161228 TransBathroom
Should transgender people – that is, people who identify themselves as the sex or gender different from the one they were born as – have to use the bathrooms of the gender they were born as, or should they be allowed to use the bathrooms of their identified gender?

5. V161227 Same Sex Service
Do you think business owners who provide wedding-related services should be allowed to refuse services to same-sex couples if same-sex marriage violates their religious beliefs, or do you think business owners should be required to provide services regardless of a couple's sexual orientation?

6. V161193 Birthright
Some people have proposed that the U.S. Constitution should be changed so that the children of unauthorized immigrants do not automatically get citizenship if they are born in this country. Do you favor, oppose, or neither favor nor oppose this proposal?

7. V161204 Affirmative Action
Do you favor, oppose, or neither favor nor oppose allowing universities to increase the number of black students studying at their schools by considering race along with other factors when choosing students?

8. V161213 ISIS
Do you favor, oppose, or neither favor nor oppose the U.S. sending ground troops to fight Islamic militants, such as ISIS, in Iraq and Syria?

9. V161229 Gay protection
Do you favor or oppose laws to protect gays and lesbians against job discrimination?

10. V161233 death penalty
Do you favor or oppose the death penalty for persons convicted of murder?

11. V162295 torture
Do you favor, oppose, or neither favor nor oppose the U.S. government torturing people who are suspected of being terrorists, to try to get information?

Not considered because they were not found to be controversial:

- V161226 Parental leave
Do you favor/oppose, or neither favor nor oppose requiring employers to offer paid leave to parents of new children?

**Freira et al. 2021 - Argentinian survey**

1. Schools should reopen before the end of the academic year.
2. Non-essential public meetings should be banned until the development of a vaccine.
3. People should be allowed to leave their homes and exercise at least once a day.
4. People over 70 years old should not be allowed to leave their homes until a vaccine is found.
5. People who are found in a public space without a valid reason should have a criminal record.
6. People should be allowed to freely travel within the country without requesting permission from the government.
7. The government should track the movements of all patients who tested positive for COVID-19 using their cell-phone data.
8. The government should fine those individuals who upload false information to the official virus-tracking app.

Not considered because they were not found to be controversial:
- The government should force the citizens to share their geolocation through an official virus-tracking app.

**Freira et al. 2021 - Brazilian survey**

1. Schools should reopen and resume face-to-face teaching by the end of the year.
2. Non-essential public meetings should be banned until the development of a vaccine.
3. Gatherings of more than 10 people should not be allowed until a vaccine is developed.
4. People over 70 years old should not be allowed to leave their homes until a vaccine is found.
5. People who are found in a public space without a valid reason should have a criminal record.
6. The government should track the movements of all patients who tested positive for COVID-19 using their cell-phone data.
7. The government should fine those individuals who upload false information to the official virustracking app.
8. The government should force the citizens to share their geolocation through an official virus-tracking app.

Not considered because they were not found to be controversial:
- All businesses and stores should reopen without requiring them to obtain an official authorization.

**Freira et al. 2021 - Uruguayan survey**

1. Universities should reopen and resume face-to-face teaching by the end of the year.
2. Gatherings of more than 10 people should not be allowed until a vaccine is developed.
3. The government should fine people who do not respect social distance in the street.
4. People diagnosed with COVID19 should have a criminal record if they are found in a public space during the period when transmission risk is high.
5. People should be allowed to freely travel within the country without requesting permission from the government.
6. The government should track the movements of all patients who tested positive for COVID-19 using their cell-phone data.
7. The government should fine those individuals who upload false information to the official virus-tracking app.
8. The government should force the citizens to share their geolocation through an official virus-tracking app.

Not considered because they were not found to be controversial:

- People over 70 years old should not be allowed to leave their homes until a vaccine is found.

**Freira et al. 2021 - American survey**

1. All schools in the United States should reopen before the end of 2020.
2. All non-essential public events should be banned until a vaccine is found.
3. The Federal Government should track the location of people infected with COVID-19 using a mobile phone app.
4. Wearing a mask in public spaces should be optional.
5. People over 70 years old should not be allowed to leave their homes until a vaccine is found.
6. People should request permission to the Federal Government to travel from one state to another.
7. Until a vaccine is found, the Federal Government should not allow mass protests in the United States.

**Pew Research 2020 - All opinion questions**

Is your overall opinion of [INSERT ITEM] very favorable, mostly favorable, mostly unfavorable, or very unfavorable?
1. The Department of Homeland Security
2. The Internal Revenue Service, the IRS
3. The Justice Department
4. The Department of Veterans Affairs, the VA
5. The Immigration and Customs Enforcement, known as ICE [PRONOUNCED: 'ice']

Not considered because they were not found to be controversial:

- The Centers for Disease Control and Prevention, the CDC
- The Department of Health and Human Services, the HHS
- The Census Bureau
- The Postal Service
- The Federal Reserve

**Pew Research 2014 - Non-political issues**

1. In the past week, did you donate money, time or goods to help the poor and needy?
2. In the past week, did you tell a white lie?
3. In the past week, did you lose your temper?
4. In the past week, did you ever eat too much?
5. In the past week, did you meditate to cope with stress?